\documentclass[prb,superscriptaddress,showpacs,preprint,preprintnumbers]{revtex4}
\usepackage{graphicx}

\newcommand{\fref}[1]{Fig.~\ref{#1}}

\newcommand{\PZT}{Pb(Zr$_{0.2}$Ti$_{0.8}$)O$_3$~}

\begin{document}
\title{Thermal quench effects on ferroelectric domain walls}
\author{P. Paruch}
\affiliation{DPMC-MaNEP, Universit\'e de Gen\`eve, 24 Quai Ernest Ansermet,
1211 Geneva, Switzerland.} \email{patrycja.paruch@unige.ch}
\author{A. B. Kolton}
\affiliation{CONICET, Centro At\'omico Bariloche, 8400 San Carlos de Bariloche,
R\'{i}o 
Negro, Argentina}
\author{X. Hong}
\altaffiliation{present address: DPA, University of Nebraska-Lincoln, Lincoln, Nebraska 68588, USA} \affiliation{DAP, Yale University, PO Box 208284, New Haven, CT 06520-0284, USA}
\author{C. H. Ahn}
\affiliation{DAP, Yale University, PO Box 208284, New Haven, CT 06520-0284, USA}
\author{T. Giamarchi}
\affiliation{DPMC-MaNEP, Universit\'e de Gen\`eve, 24 Quai Ernest Ansermet,
1211 Geneva, Switzerland.}
\date{\today}

\begin{abstract}
Using piezoresponse force microscopy on epitaxial ferroelectric thin films, we
have measured the evolution of domain wall roughening as a result of heat-quench
cycles up to 735$^\circ$, with the effective roughness exponent $\zeta$ changing
from 0.25 to 0.5. We discuss two possible mechanisms for the observed $\zeta$ increase: a quench from a thermal 1-dimensional configuration, and from a locally-equilibrated pinned configuration
with a crossover from a 2- to 1-dimensional regime. We find that the post-quench spatial structure of the metastable states, qualitatively consistent with the existence of a growing dynamical length scale whose ultra slow evolution is primarily controlled by the defect configuration and heating process parameters, makes the second scenario more plausible. This interpretation suggests that pinning is relevant in a wide 
range of temperatures, and in particular, that purely thermal domain wall configurations might not be observable in this  glassy system. We also demonstrate the crucial effects of oxygen vacancies in stabilizing domain structures.
\end{abstract}

\maketitle

\section{Introduction}
Domain walls separating differently polarized regions in ferroelectric thin
films present a powerful model system in which the characteristic roughening and
complex dynamics resulting from competing elastic and pinning forces can be
readily accessed  \cite{paruch_dw_review_07}. Understanding this behavior is key
to the physics of many diverse disordered elastic systems (DES) \cite{agoritsas_physb_12_DES,yamanouchi_prl_06_creep_ferromagnetic_semiconductor,ponson_prl_06_fracture,lemerle_prl_98_FMDW_creep,blatter_94_4mp_vortex_review,giamarchi_varenna_wigner_review,gruner_rmp_88_CDW_review,kardar_review_lines}  and of significant technological interest for memory and electromechanical ferroelectric
applications \cite{scott_memories,waser_memories,kumar_apl_04_SAW}, and devices using the
domain walls as nanoscale functional components
\cite{bea_natmat_09_domainwalls_NV,salje_cpc_10_multiferroic_boundaries}, in
which control over the stability and growth of domain structures is of paramount
importance.

Theoretically, although equilibrium DES properties are relatively well
understood, much less is know about the out-of-equilibrium behavior of these
systems. A major challenge is understanding the non-steady slow dynamics
associated with aging \cite{cugliandolo_98_aging_review,cugliandolo_prl_06_keldysh_elastic,schehr_frg}. A specially interesting realization of such out-of-equilibrium phenomena is provided by quenches, in
which a parameter is abruptly varied. The study of such quenches is a challenge,
not only for classical systems but also for quantum systems
\cite{cugliandolo_prb_02_dynamics_quantum_aging,polkovnikov_rmp_11_quantum_nonequilibrium}. For classical systems, studies of interfaces subjected to quenches 
\cite{kolton_prl_05_flat_interface,bustingorry_prb_07_vortex_glass,iguain_prb_09_aging_line,kolton_prb_06_slowrelaxatdepinning,kolton_prl_09_agingatdepinning} 
have shown aging of the interface and a long term memory of the initial configuration.  

On the experimental side, studies of ferroelectric domain walls in the DES
framework have so far focused mostly on equilibrium properties, with switching
current and dielectric measurements showing domain wall pinning in bulk
ferroelectrics \cite{damjanovic_prb_97_piezo}, and random-field disorder
inferred for relaxors \cite{kleeman_ferro_98_dw_rf}, while piezoforce microscopy
(PFM) studies of domain wall roughening and creep reported strong dipolar
interactions and random bond disorder in epitaxial ferroelectric thin films
\cite{tybell_prl_02_creep,paruch_prl_05_dw_roughness_FE,paruch_jap_06_dynamics_FE} . Unlike their
theoretical counterparts, these systems also show pinning by individual strong
defects \cite{rodriguez_apl_08_DW_defect}, and local variations of both disorder
strength and universality class \cite{jesse_natmat_08_SSPFM}, further
complicating the approach to non-equilibrium behavior.  Thermal effects in bulk
\cite{likodimos_prb_02_thermal_kinetics_TGS} and thin film materials
\cite{paruch_apl_06_stability} include increased roughness and higher domain
wall mobility upon heating, and transition to domain wall depinning at lower
temperatures \cite{jo_PRL_09_dw}. However, a detailed nanoscale analysis of the
complex behavior associated with a thermal quench of these systems remains an
open question.

In this paper we report on PFM measurements of domain walls in ferroelectric
\PZT (PZT) thin films, heat-quench cycled up to 735$^\circ$C, in which the value
of the roughness  exponent $\zeta$, reflecting the system dimensionality and the nature of the disorder, is extracted from an analysis of the domain wall
position, and evolves from $\sim$0.25 to $\sim$0.5. Despite the similarity of the
measured exponent with the thermal $\zeta_{th} = 1/2$ in one dimension, we show
that the observed behavior cannot result from the quench of an initial high
temperature thermal configuration, as the observed roughness values are much
higher than would be expected in this scenario. Rather, it is clear that
disorder pinning continues to dominate the behavior during the quench, with one
possibility being that the domain walls cross from a 2- to 1-dimensional regime.
We also follow the complex thermal evolution of circular nanoscale ferroelectric
domains, in which the combined effects of disorder pinning, line tension and a
preferential polarization orientation can be observed, and the crucial effects
of oxygen vacancies in stabilizing domain structures are highlighted.

\section{Materials and Methods}
Thermal quench studies were carried out on {\it c}-axis oriented PZT films,
epitaxially grown on single crystal (001) Nb:SrTiO$_3$ substrates at
$\sim$500$^{\circ}$C by off-axis radio-frequency magnetron sputtering, in an
Ar-O$_2$ process gas mixture \endnote{Ar:O$_{2}$ = 58:42, 180 mTorr and
Ar:O$_{2}$ = 80:20, 225 mTorr were used by P. P. and X. H.,  respectively}, 
with high crystal and surface quality confirmed by x-ray diffraction and atomic
force microscopy (AFM). In these films, the polarization vector along the c-axis
can be locally switched by a biased AFM tip, using the substrate as a ground
electrode, and the resulting ferroelectric domains imaged by PFM
\cite{kalinin_mm_06_PFM_review}. To study the effects of heat-quench cycling on
domain wall roughness we wrote 10 $\times$ 10 $\mu$m$^2$ arrays of linear 1
$\mu$m wide domains with $\pm12$ V alternately applied to a scanning AFM tip. To
explore the interaction between heating, line tension and disorder pinning, we
wrote arrays of nanoscale circular domains with $\pm12$ V pulses of different
duration in a uniform, oppositely polarized region. The films were heated in air
to progressively higher temperatures for fixed time intervals, then quenched to
room temperature on a large copper block. Ambient temperature PFM images of the
evolving domain structures were acquired after each heating-quench cycle.  In
addition, to investigate the role of oxygen vacancies on domain stability, films
grown under the same conditions but cooled in either process gas, Ar, or O$_2$
were compared.

As in our previous work \cite{paruch_prl_05_dw_roughness_FE}, to extract the
correlation function of relative displacements $B(L)$ for a given PFM phase
image of linear domains we first isolated a single domain wall, then binarized
the image using a cutoff extracted from the midway point between the phase
contrast values for the two polarization orientations.  The elastically optimal
flat domain wall position was determined from a least-squares linear fit to the
binarized image, which finally allowed the extraction of the relative
displacements, and their correlation at different length scales. To compensate
for possible effects of scanner drift, for each set of walls PFM measurements taken in an
``up'' vs ``down'' scan direction were averaged. Although ambient-condition
walls after writing were generally well defined, repeated heat-quench cycles, especially close to $T_C$, led to the appearance of multivalued features such as bubbles or overhangs.  In some cases, such features could also be present even immediately after writing at room temperature. Bubbles, being very small, well defined domains of opposite contrast near but not coincident with the position of the primary domain wall,
could be removed from the binarized image by hand.  For overhangs, the algorithm
used to calculate relative displacements would simply result in a renormalized
single value ``effective position'' of the domain wall.  Therefore, when such
features became prevalent, or when the films switched to the preferred
polarization direction upon heating near $T_C$, quantitative analysis could no
longer be carried out, as indicated by the symbol ``x'' in Table.~\ref{zeta_values}. 

\section{Domain wall roughness}
Generally, an elastic manifold in the absence of disorder and thermal fluctuations (0K temperature) would take an ideally flat configuration in order to optimize its energy.  Random variations in the potential landscape, whether due to disorder or thermal fluctuations, allow further optimization as the manifold wanders between particularly favorable regions in the potential, giving rise to a characteristic roughening. In the case of single manifold, an interface such as a domain wall, contact line, or fracture, fully in equilibrium with its random potential landscape, the resulting geometrical fluctuations from an elastically optimal flat configuration are present at all length scales, with the system showing mono-affine scaling properties (for a more detailed presentation, see \onlinecite{agoritsas_physb_12_DES} and references therein). The key quantity describing this roughening is the correlation function:
\begin{equation}\label{eq:B_gen}
 B(L)  = \overline{\langle [u(z+L)-u(z)]^2 \rangle}  \propto L^{2\zeta}\ \ ,
\end{equation}
known as the roughness function.  Here $u(z)$ are the transverse displacements along the longitudinal coordinate $z$ with respect to a flat configuration, measured a distance $L$ apart on the interface, and $\langle ... \rangle$ and $\overline{...}$ the thermal and disorder averages, respectively. The value of the roughness exponent $\zeta$ depends on the dimensionality of the system, the nature of the disorder potential, and the range of the elastic interactions. Extracting $\zeta$ not only yields this important information, which may be used to determine the universality class of the disorder and the dominant pinning defects present in the system, but also allows a full scaling prediction of the complex non-linear creep response of such systems to small driving forces \cite{tybell_prl_02_creep}.
We have previously shown for linear domains created artificially by straight-line scanning with a biased AFM tip that, as a result of this writing process, the domain walls are essentially flat at large length scales, with roughening observed only at short length scales, probably where the stray fields during the writing process allow some small accommodation to the potential landscape \cite{paruch_prl_05_dw_roughness_FE}.  The experimentally obtained $B(L)$ thus shows a power-law growth at short length scales, from which the roughness exponent $\zeta$ can be extracted, followed by a saturation at large length length scales.

\begin{figure}
\includegraphics[width=0.5\textwidth]{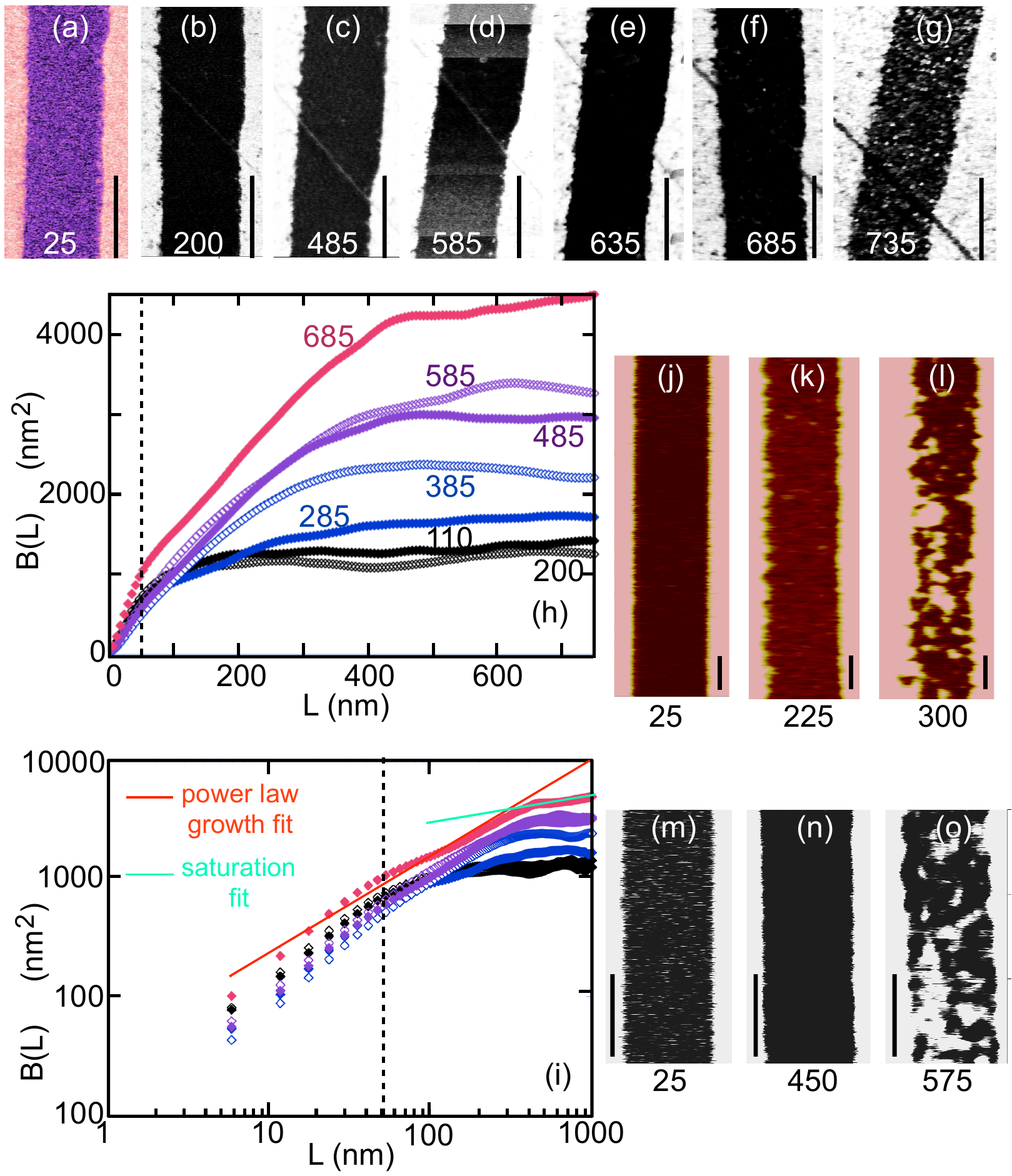}
\caption{PFM images of linear ferroelectric domains in a process-gas-cooled 50
nm PZT
film (a) immediately after writing and (b--g) after thermal cycling to the
indicated temperatures ($^\circ$C).  Similar measurements (j--l) in an
O$_2$-cooled 100 nm film and (m--o) an Ar-cooled 40 nm film, where higher oxygen
vacancy densities appear to stabilize the domain walls. Vertical 1 $\mu$m scale
bars are shown in all images. (h) Average $B(L)  = \overline{\langle
[u(z+L)-u(z)]^2 \rangle}$ extracted after thermal cycling to the indicated
temperatures ($^\circ$C) for all the domain walls in the 50 nm film. (i) The power-law growth region of $B(L)$ for same measurement (identical color code for the temperatures). The extrapolated fit to the power-law growth region of $B(L)$ (first 20 points) with $\ln[B(L)] \propto 0.890\ln[L]$ is shown in red, and the fit to the saturation region of $B(L)$ in aquamarine for the 685$^\circ$ data set.  The intersection of these fits allows the saturation length scale $L^\star$ to be extracted. The dashed vertical line in (h) and (i) indicates the length scale $L$ equal to the film thickness. }
\label{straight_DW}
\end{figure}

As previously reported in \cite{paruch_apl_06_stability}, linear domains in
process-gas-cooled films show very high stability upon heating, with increased
roughness only at small length scales  (\fref{straight_DW}(a--f)).  At
625--785$^\circ$C (depending on the sample), close to the Curie temperature $T_C$ of the films determined
by x-ray diffraction \cite{gariglio_apl_07_PZT_highTc}, extensive polarization
switching occurs and the written domains eventually disappear as the region
reverts to its as-grown monodomain state. Quasi-identical behaviour was observed
in the Ar-cooled samples (\fref{straight_DW}(m--o)). In the O$_2$-cooled films
however, the domain structures showed much lower thermal stability.
In these films, expected to show decreased oxygen vacancy densities compared to process-gas or Ar-cooled samples, increased roughening was found even for short (20 min.) heating
intervals, with domain disappearance observed already at 250--350$^\circ$C (\fref{straight_DW}(j--l)). Nonetheless, x-ray measurements confirmed sample $T_C$ in the 600-700$^\circ$C range, similar
to those of the process-gas-cooled samples \cite{gariglio_apl_07_PZT_highTc}.

To quantitatively examine the roughness evolution, for each sample we extracted
the average roughness function $B(L)$. For the relatively flat, as-written domain walls before thermal cycling, as in our previous studies \cite{paruch_prl_05_dw_roughness_FE}, we find a power-law growth of $B(L)$ at short length
scales followed by saturation at  $L^\star \sim$25--100 nm (comparable to the $\sim$40--100 nm film thickness), with $B(L^\star)\sim$100--1000 nm$^2$ in the different films. After heating, the power-law-growth region
of $B(L)$ extends to $L^\star \sim$300--700 nm and the saturation value increases ($B(L^\star) \sim$1000--10000
nm$^2$), consistent with the observed increase in domain wall roughness
(\fref{straight_DW}(h,i)). From the power-law growth region of $B(L)$ (\fref{straight_DW}(l)), we extract
a value for the roughness exponent $\zeta$, characterizing the static
equilibrium configuration of the domain wall in the random manifold regime,
where $B(L) \propto L^{2\zeta}$. As shown in Table.~\ref{zeta_values} for domain
walls in different films, heating increases the value of $\zeta$ from $\sim$0.25
at ambient conditions and lower temperature thermal cycling to $\sim$0.5 at
higher temperatures and near domain disappearance \endnote{(A) The 50 nm film
was heated for 3 hour intervals in 100$^\circ$C steps to 485$^\circ$C, then in
50$^\circ$C steps till domain disappearance at $\sim$785$^\circ$C. The 52 and 91
nm films were heated for 1 hour intervals in 100$^\circ$C steps to 485$^\circ$C,
then in 25$^\circ$C steps till domain disappearance at $\sim$625$^\circ$C. (B)
The 50 nm film was heated for 30 minute intervals in 100$^\circ$C steps to
300$^\circ$C, 50$^\circ$C steps to 450$^\circ$C, then 25$^\circ$C steps till
domain disappearance at $\sim$525$^\circ$C. The 40 nm films were heated for 20
minute intervals in 100$^\circ$C steps till 400$^\circ$C, 50$^\circ$C steps till
550$^\circ$C, then 25$^\circ$C steps till domain disappearance at
$\sim$625$^\circ$C. The 100 nm film was heated for 20 minute intervals in
100$^\circ$C steps till 200$^\circ$C, then 25$^\circ$C steps till domain
disappearance at $\sim$325$^\circ$C.}. In O$_2$-cooled films, higher roughness and increased $\zeta$
values are seen already at relatively low temperatures.
\begin{table}[h]
\caption{Thermal evolution of the roughness exponent $\zeta$ for PZT films
cooled in process gas (PG), argon (Ar), and oxygen (O$_2$) after long (A: 1--3
hrs., P.P.) and short (B: 20--30 min., X.H.) heat-quench cycles. - data set
missing or noisy. x  temperatures beyond domain disappearance, multivalued wall roughness configuration. * 225$^\circ$C
measurement.}\label{zeta_values} \centerline{
\begin{tabular}{|c|c|c|c|c|c|c|c|c|}
\hline A) $T$($^\circ$C) & 25 & 110 & 200 & 285 & 385 & 485 & 585 & 685 \\
\hline 50 nm PG & 0.33 & 0.43 & 0.42 & 0.48 & 0.54 & 0.50 & 0.51 & 0.45 \\
\hline 91 nm PG & 0.23 & 0.22 & 0.35 & - & 0.37 & 0.39 & x & x \\
\hline 52 nm PG & 0.26 & 0.26 & 0.18 & 0.23 & 0.26 & 0.31 & 0.45 & x \\
\hline
\hline
\hline B) $T$($^\circ$C) & 25 & 200 & 300 & 400 & 475 & 500 & 575 & 600\\
\hline 50 nm PG & 0.10 & 0.15 & 0.31 & 0.30 & 0.27 & 0.42 & x & x\\
\hline 40 nm PG & 0.24 & - & 0.29 & 0.29 & - & 0.24 & 0.42 & 0.50 \\
\hline 40 nm Ar & 0.27 & 0.18 & 0.17 & 0.27 & - & 0.22 & 0.38 & 0.41 \\
\hline 100 nm O$_2$ & 0.15 & 0.45* & 0.47 & x & x & x & x & x \\
\hline
\end{tabular}}
\end{table}

\section{Discussion}
To understand these data, we first consider the thermal roughening of the linear
domain walls within the theoretical framework of DES. Our previous studies have
shown that at ambient conditions with no applied electric field, thermal excitations alone are not enough to overcome the energy barriers between different metastable states in the duration of the
experiment, although the electric field applied during writing allows domain
walls to accommodate at short length scales to the surrounding random potential
landscape.  A power-law growth of $B(L)$ is thus observed only at length scales
smaller than the film thickness, where the domain wall behavior agrees well
with theoretical predictions for 2-dimensional interfaces pinned by weak,
collective random bond disorder and in the presence long-range dipolar forces
\cite{paruch_prl_05_dw_roughness_FE,nattermann_jopc_83_dipole_disorder}.  Given
the finite thickness of the sample, much smaller than its lateral dimensions, it
is reasonable to expect that in a weak collective pinning scenario, were the
walls to fully equilibrate, 1-dimensional behavior would be evident at large
length scales. Although it is difficult to estimate the exact length scale $L_{\times}$ for such a
2-dimensional to 1-dimensional crossover in a disordered system, we can expect it to be of the order of
the sample thickness. This would be exactly the case for purely thermal roughening.
The increased energy provided to the samples on thermal cycling to progressively higher
temperatures is one way to promote further equilibration of the domain walls,
allowing them to accommodate to the surrounding potential landscape at length
scales greater than the film thickness, where the 1-dimensional nature of the domain walls should become evident.

\begin{figure}
\centerline{\includegraphics[width=0.5\textwidth]{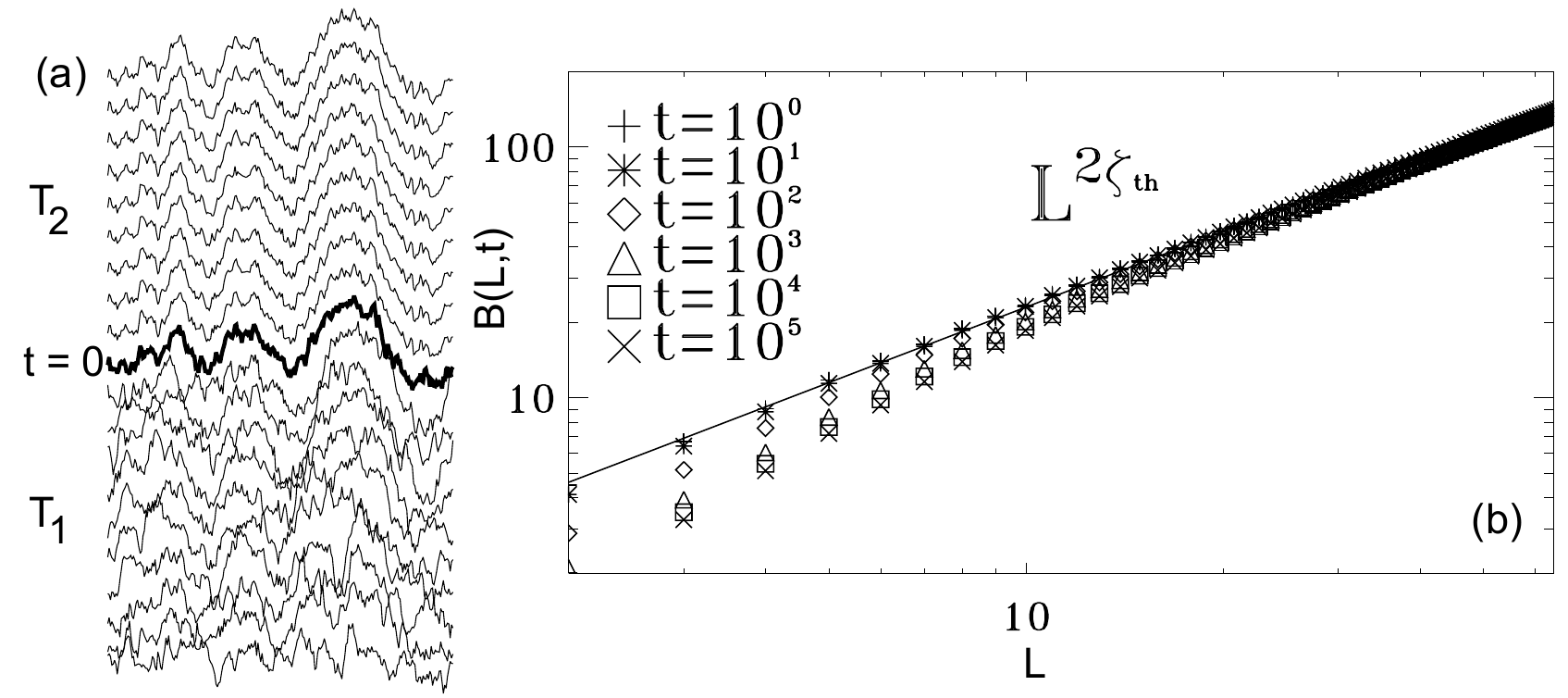}}
\caption{Numerical simulation of an overdamped elastic string in a random medium
undergoing a temperature quench $T_1 \rightarrow T_2 < T_1$ at $t=0$.
(a)  From bottom to top: Typical evolution of the elastic string each $10000$ steps. 
The bold line is the configuration at the quench. (b) 
From top to bottom curve: Evolution of the
disorder-averaged correlation function of relative displacements
$B(L,t)$ after the quench (symbols). The roughness exponent
$\zeta_{th}=0.5$ describes the geometry of the string at large
length scales (solid line).} \label{fig:molsim}
\end{figure}

If thermal roughening dominates in the high temperature 1-dimensional regime,
the walls should be characterized by a roughness exponent $\zeta_{th} = 1/2$,
and slowly age to the equilibrium $\zeta_{eq}$ configuration when quenched. For
an increasing time $t$ after the quench, a growing length $L_{dyn}(t)$ should thus
separate the equilibrated short length scales (with exponent $\zeta_{eq}$) from the large
scales conserving a memory of the roughness at the quench (with exponent $\zeta_{th}$),
analogous to a quench from an initially flat configuration
\cite{kolton_prl_05_flat_interface}. We confirmed such a scenario in Langevin
dynamics simulations of a one-dimensional interface in a random medium as
detailed in Ref.\onlinecite{kolton_prl_05_flat_interface}, but this time quenching
an initially equilibrated high-temperature configuration. The results indeed
show the quasi-freezing of this initial configuration at the largest length scales,
with thermally activated dynamic evolution distinguishable only at very small
length scales (\fref{fig:molsim}(a)). This behaviour is also reflected in the
evolution of $B(L,t)$ after the quench (\fref{fig:molsim}(b) ): at large lengths
$L>L_{dyn}(t)$ we
observe $B(L,t)\sim L^{2\zeta_{th}}$, while at small lengths the
line tends to approach the equilibrium roughness exponent
$\zeta_{eq}$ (with $\zeta_{eq}=2/3$ for an elastic line with short
range elasticity (random bond disorder)).

If the barriers are large, as the high stability of ferroelectric domain walls
indicates, the length $L_{dyn}(t)$ can grow extremely slowly, and most of the
observable interface could well be characterized by the exponent $\zeta_{th} =
0.5$. This scenario seems consistent with the exponent $\zeta \sim 0.5$ observed
at length scales up to 10 times greater than the thickness of the film. However,
if we consider an elastic 1-dimensional interface with energy $H =
\frac{c}{2}\int dz(\nabla u)^2$, we obtain its thermal roughening at temperature
$T$ as 
\begin{equation}
B(L) = \frac{k_BT}{c}\int^{+\infty}_{-\infty}dk\frac{1-\cos(kL)}{2\pi k^2} =
\frac{k_BT}{2c}|L|
\end{equation}
where the elastic constant $c$ in the case of the domain wall is the product of
its energy per unit area $\sigma_{DW}$ and the film thickness $d_f$ and $k_B$ is
the Boltzmann constant. Taking $\sigma_{DW} =132$ mJm$^{-2}$ from ab-initio
calculations for PbTiO$_3$ \cite{meyer_prl_02_PTO_DW}, we can therefore
calculate the gradient $S = \frac{k_BT}{2c} = \frac{k_BT}{2\sigma_{DW}d_f}$, and the roughness $B(L^\star)_{th}$ expected for the thermal roughening scenario at the length scale
corresponding to the experimentally observed saturation length $L^\star$. These
values can be directly compared to the experimentally obtained values for the roughness $B(L^\star)_{exp}$ and the gradient $S_{exp} =
B(L^\star)_{exp}/L^\star$ in the power-law-growth region where $B(L) \propto
L^{2\zeta}$. As shown in Table.~\ref{slopes} for the 50 nm PZT film of Fig. 1,
the experimental $B(L)$ values are many orders of magnitude higher than those expected for a
purely thermal roughening of a 1-dimensional interface. The observed behavior
at large length scales therefore cannot be simply a quench of the high
temperature configuration. Rather, pinning by disorder in the film continues to
be the dominant effect, even during the quench.
\begin{table}[h]
\caption{Comparison of the roughness function $B(L)$ and its gradient $S=
\frac{k_BT}{2c}$ for a thermally roughened 1-dimensional interface with the
experimentally obtained gradient of $B(L) \propto L^{2\zeta}$ in a 50 nm PZT
film.}\label{slopes} \centerline{
\begin{tabular}{|c|c|c|c|c|c|c|c|}
\hline $T$ (K) & 383 & 473 & 558 & 658 & 758 & 858 & 958 \\
\hline $S \times 10^{-4}$ (nm) & 4.0 & 4.9 & 5.8 & 6.9 & 7.9 & 9 & 10 \\
\hline $L^\star$ (nm) & 58 & 121 & 72 & 180 & 199 & 200 & 232 \\
\hline $B(L^\star)_{th}$ (nm$^2$) & 0.023 & 0.060 & 0.042 & 0.124 & 0.158 & 0.179 & 0.232
\\
\hline $B(L^\star)_{exp}$ (nm$^2$) & 275 & 1593 & 1143 & 2805 & 3065 & 4089 &
4343 \\
\hline $S_{exp}$ (nm) & 4.8 & 13.1 & 15.8 & 15.5 & 15.4 & 20.5 & 188 \\
\hline
\end{tabular}}
\end{table}

\begin{figure}
\centerline{\includegraphics[width=0.5\textwidth]{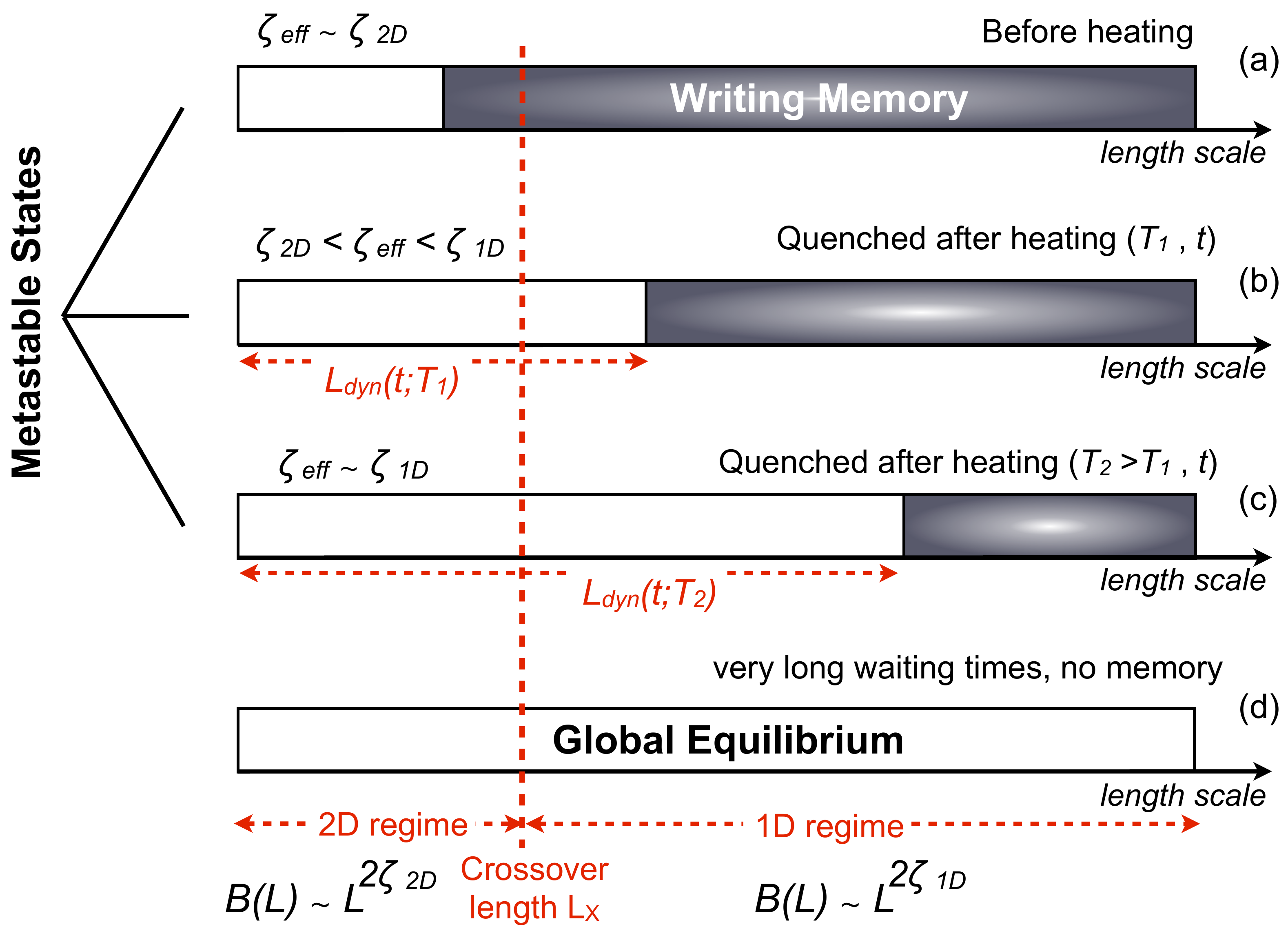}}
\caption{Disorder dominated scenario as a function of length scale.
(a) At ambient temperature before heating, the as-written, initially flat  domain walls are only able to locally equilibrate at short  length-scales. (b) and (c) Heating facilitates thermally activated 
glassy relaxation and longer length scales become locally equilibrated 
as the heating temperature (or heating time) increases. 
(d) If fully in equilibrium (reference state), the system displays a 2-dimensional 
to 1-dimensional crossover at length scale $L_{\times}$, with no memory of the written state.  This may be reflected by the effective roughness exponent $\zeta_{eff}$ in the correlation 
function $B(L)\sim L^{2\zeta_{eff}}$ of the quenched metastable states,  
as in (b) and (c), evolving from the 2-dimensional value $\zeta_{1D}$, when $L_{dyn}\leq L_{\times}$,
towards the 1-dimensional value $\zeta_{1D}$ when $L_{dyn} \gg L_{\times}$, especially if the crossover length scale is smeared out. In the 1d regime the ``dynamical length'' $L_{dyn}(t;T)$ is 
expected~\cite{kolton_prl_05_flat_interface} to grow as 
$L_{dyn}(t;T)\sim [(T/U_0) \log(t/t_0)]^{1/\theta}$, with $\theta$ a positive exponent, $U_0$ a characteristic 
temperature, and $t_0$ a characteristic microscopic time.
This prediction agrees qualitatively with the experimentally observed increase in $\zeta$ and $L^\star$ values with heating time and temperature. 
} \label{fig:scenariodisorder}
\end{figure}

One possible scenario is that of a dimensional crossover dominated by disorder:
in equilibrium with a random bond disorder potential, below a typical 
length scale $L_{\times}$, comparable to the film thickness, the domain walls should act as 2-dimensional
elastic sheets with $\zeta_{2D} \simeq 0.2$--0.3, and as 1-dimensional elastic
strings with $\zeta_{1D} = 2/3$ for higher length scales 
(see Fig.\ref{fig:scenariodisorder}(d)). If this crossover is
smeared out, it is not improbable that an effective $\zeta_{\times} :
\zeta_{2D} <  \zeta_{\times} < \zeta_{1D} $ is observed experimentally. 
According to the theory of DES \cite{kolton_prl_05_flat_interface}, 
the locally equilibrated regime just after the quench should 
extend up to a length-scale that grows  
with the heating-time $t$ and the heating temperature $T$,
as the dynamical length $L_{dyn}(t)\sim[(T/U_0) \log(t/t_0)]^{1/\theta}$, 
with  $\theta$ a positive exponent, $U_0$ a characteristic 
temperature, and $t_0$ a characteristic microscopic time. 
We then expect to observe equilibrium power law behavior 
(possibly including the equilibrium dimensional crossover) 
below $L_{dyn}(t)$ and a memory of the flat initial 
condition, imposed by the linearly scanning AFM tip during writing, above it. The theory thus predicts a temperature dependence of the effective roughness exponent, 
from $\zeta_{\times} \sim \zeta_{2D}$ at low 
temperatures, $T: L_{dyn}(t;T) \sim L_{\times}$ 
(see Fig.\ref{fig:scenariodisorder}(b)), to 
$\zeta_{\times} \sim \zeta_{1D}$ at higher temperatures 
$T: L_{dyn}(t;T) \gg L_{\times}$ (see Fig.\ref{fig:scenariodisorder}(c))
where the power law behavior is effectively dominated by the equilibrium 1-dimensional regime.
The results displayed Fig.\ref{straight_DW}(e,l) qualitatively agree with this theoretically expected scenario,
and in particular, the results in table~\ref{zeta_values} are consistent 
with the expected heating-temperature dependence of the effective 
roughness exponent. The theory also predicts, at a fixed temperature, 
a slow (logarithmic) time dependence of $L_{dyn}(t)$ that should translate into a 
heating-time dependence of $\zeta_{\times}$, from 
$\zeta_{2D}$ at short heating times $t: L_{dyn}(t;T) \sim L_{\times}$, 
to $\zeta_{1D}$ at large 
times, $t: L_{dyn}(t) \gg L_{\times}$, as summarized in Fig.\ref{fig:scenariodisorder} .
In this respect, the results shown in Table~\ref{zeta_values} for 
the 50 nm PG-cooled samples in (A) and (B) seem to be also qualitatively 
consistent with the theory. 

Moreover, the $S_{exp}$ values in table~\ref{slopes} show no clear linear temperature dependence, a 
fact that is also qualitatively consistent with large length scales dominated 
by disorder, and which we can further analyze through the following estimates.  
Phenomenologically, in the case of disorder-dominated roughening of an elastic interface \cite{larkin_70,agoritsas_physb_12_DES} , we can
consider in the 2-dimensional regime at short length scales that $B(L) = r_f^2(L/L_{C})^{2\zeta(2D)}$, where $r_f$ is the correlation length of the disorder potential and $L_C$ the Larkin length (minimum pinning length, below which the interface would behave purely elastically). In the 1-dimensional regime at large length scales, this relation would become $B(L) = r_{char}^2(L/L_{\times})^{2\zeta(1D)}$, where the characteristic length in this case would be the crossover length scale $L_{\times}$ and $r_{char} = r_f (L_{\times}/L_C)^{\zeta(2D)}$.
Taking $L_C \sim 0.2$ nm from \onlinecite{paruch_jap_06_dynamics_FE}, $\zeta(2D) = 0.25$, $\zeta(1D) = 0.5$, and $L_{\times}\sim d_f$ for the 50 and 91 nm samples, we find that the experimentally obtained $B(L)$ values would correspond to $r_f \sim 4$ nm in the 2-dimensional regime (ambient  conditions) and $r_f \sim 4$--8 nm in the
1-dimensional regime (thermally cycled).  For reference, the  $10^{18}$--$10^{19}$cm$^{-3}$ impurity
densities extracted from current-voltage measurements in PZT films 
\cite{zubko_jap_06_PZT} correspond to an inter-defect spacing of $\sim
4.5$--10 nm, and would thus qualitatively agree with the extracted disorder correlation lengths. 

An additional important consideration, especially at increasing temperatures, is the exact nature of the disorder. Although surface deterioration is seen upon heating, especially at the highest temperatures, x-ray characterization of the samples demonstrates continued high crystalline quality
\cite{paruch_apl_06_stability}.  However, the mobility of point defects, such as oxygen vacancies, increases strongly with temperature. From the comparison of PG-, Ar- and O$_2$-cooled films, it appears
that the primary defects responsible for the domain wall pinning are indeed oxygen
vacancies. These defects can appreciably lower their energy by associating with a domain wall \cite{he_prb_03_DW_ovac}, so the probability of a migration and segregation of defects at the domain walls upon thermal cycling, effectively increasing the strength of the pinning
potential and stabilizing the domain structures against large-scale thermal
roughening, is very high.   In this respect it is worth noting 
that in the theory of DES an analogous ``slow structural relaxation'' mechanism was very recently shown \onlinecite{jagla_jgr_10_pinning_relaxation} to provide a minimal model for predicting pinning enhancement 
or aging of the static friction force, and the phenomenon of ``aftershocks''  or static-relaxation-induced instabilities in the interface. This is interesting as these phenomena are both absent 
in the standard DES models with truly quenched disorder, but they are for instance observed in solid friction experiments. 

Another possibility is that the thermal cycling steps, by removing the water layer
normally present on the film surface in air at ambient temperature, significantly
affect the electrostatic boundary conditions, and thus the effective potential
landscape experienced by the domain walls.  In this case, the higher roughness
and increased $\zeta$ values could reflect the configuration adapted by the
walls under these new conditions, and frozen-in after a quench from high
temperature, possibly in parallel to a dimensional crossover. Further investigation would be needed to asses the potential importance of this effect and also to quantitatively test the scenario 
described above, summarized in Fig.\ref{fig:scenariodisorder}.

\section{Domain switching and growth}
\begin{figure}
\includegraphics[width=0.5\textwidth]{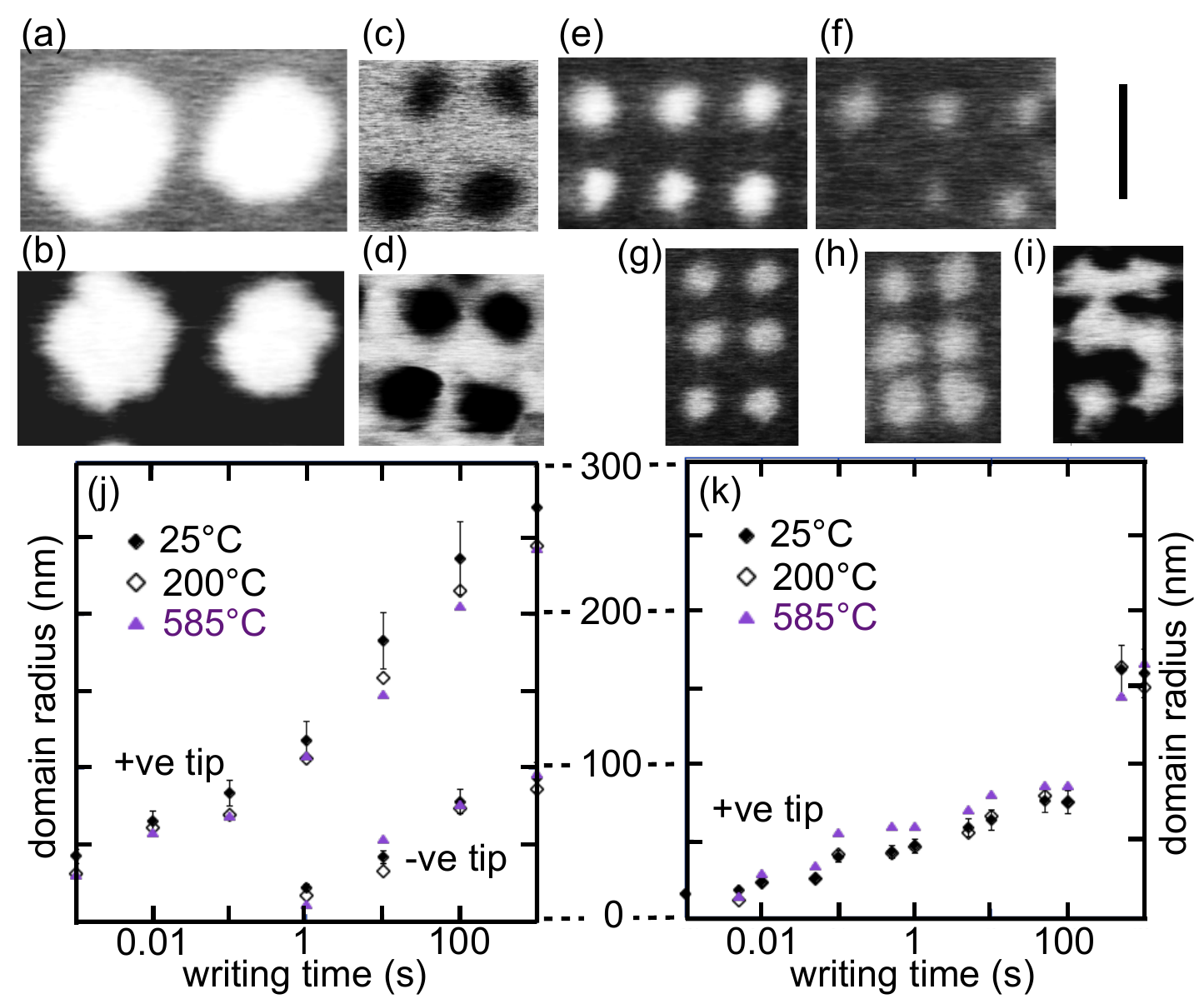}
\caption{PFM images of nanoscale ferroelectric domains written at ambient
conditions in the 91 nm PZT film ($\rm P_{UP}$ as grown) with (a) 1 s, +12 V,
(c) 10 s, -10 V, (e) 1 ms, +12 V pulses, and in a 52 nm PZT film ($\rm P_{DOWN}$
as grown) with (g) 100 ms, +12 V pulses. Domain evolution after thermal cycling
(f,h) to 200$^\circ$C and (b,d,i) 585$^\circ$C. Vertical bar $= 250$ nm. All
images are shown to the same scale. The average domain size as a function of the
writing time at ambient conditions, and after thermal cycling up to the
indicated temperatures (j) for the 91 nm film and (k) for the 52 nm
film.}\label{circular_DW}
\end{figure}

The crucial role of disorder in stabilizing the domain walls in a given
configuration is further highlighted by the behavior of nanoscale circular
domains followed in parallel measurements during the same heat-quench cycles. In these
measurements, we note the strong polarization switching asymmetry: $\rm
P_{DOWN}$ domains written with positive tip voltage in a negatively
pre-polarized area grow significantly larger for a given pulse duration 
(\fref{circular_DW}(a-f)), even in films $\rm P_{UP}$ polarized as-grown.  We
relate this feature to the highly asymmetric device configuration
\cite{maksymovych_nl_11_BFO_DW_conductivity,guyonnet_am_11_DW_conduction}.  In
PG- and Ar-cooled films at ambient temperature, both tip voltage polarities give
uniform, circular domains, which become irregular upon heating, with apparent
bowing of the domain walls. In $\rm P_{UP}$ as-grown films, $\rm P_{DOWN}$
domains decrease in size after thermal cycling, while in $\rm P_{DOWN}$ as-grown
films a slight increase in domain size was observed. Some of the smaller domains
(20--50 nm radius) of both polarities collapse even after low-temperature
heating intervals (\fref{circular_DW}(e--f)). In $\rm P_{DOWN}$ as-grown films,
significant displacements and an ``amoeboid-like'' growth are observed for $\rm
P_{DOWN}$ domains near $T_C$, with the structures eventually coalescing
(\fref{circular_DW}(g--i)). In the O$_2$-cooled films, even at ambient
conditions $\rm P_{UP}$ domains were unstable and $\rm P_{DOWN}$ domains were
stable only for domain radii greater $\sim 100 nm$.

This complex thermal evolution can be understood by considering the forces on
the domain wall due to its curvature and the polarization orientation as well as
disorder and thermal effects.  A preferred monodomain state indicates an
asymmetry of the ferroelectric double well potential, which promotes growth or
collapse of domains depending on their polarization orientation. In addition,
the ``line tension'' or elastic energy cost of circular domain walls promotes domain collapse, its effects
diminishing with increasing domain size. Finally, disorder allows domain wall
pinning, and therefore deformations from the elastically optimal circular
geometry. The relative strength of these contributions during thermal cycling
can be gauged from the degree of domain stability. Although polarization
orientation and device asymmetry \endnote{The significant asymmetry in domain
switching and stability as a function of the polarity of the tip bias is a
feature already noted in previous studies \cite{guyonnet_am_11_DW_conduction}
and can be related to the particular device geometry used.  Essentially, the
metallic electrodes on either side of the ferroelectric thin films can be
considered as two back-to-back Schottky diodes, with the barrier height
determined by the work functions of the respective electrode materials
\cite{zubko_jap_06_PZT,maksymovych_nanotech_11_PZT_conduction}. In addition,
further asymmetry in current-voltage and switching characteristics can result
from strain and other boundary-specific conditions, and, in thin films, the
direction of polarization. Finally, the device architecture itself is strongly
asymmetric: one electrode is planar, while the other is the nanoscale AFM tip,
with a nominal radius at apex form metal-coated tips given as 25-50 nm.  In such
devices, the nucleation and growth of domains is therefore strongly dependent on
the polarity of the voltage applied to the AFM tip.} determine the domain size
for a given writing time, for domains of comparable size, line tension determines
stability. Larger $\rm P_{UP}$ and $\rm P_{DOWN}$ domains remain, while 20--50
nm radius domains readily collapse on heating. Domain wall roughening is
observed for all domain sizes, although the extent of the deformations increases
with domain size. Disorder appears to be crucial in stabilizing the domain
structures, in particular the smallest domains, and protects them against
thermal effects.

\section{Conclusion}
In conclusion, our studies show that in epitaxial ferroelectric thin films,
domain wall roughening is governed primarily by disorder pinning, even upon
heating to progressively higher temperatures, followed by a quench.  However,
the increased energy provided by the thermal cycling does allow accommodation of
the artificially straight AFM-written domain walls to the surrounding disorder
potential at progressively larger length scales, with the roughness exponent
evolution possibly related to a  2-to-1 dimensional crossover in the behavior of
the system. From a comparison of O-poor vs O-rich PZT films, it appears that
oxygen vacancies are particularly important for pinning domain walls, and allow
even relatively small domains to be stabilized at high temperatures, close to
$T_C$.  This observation is especially interesting for potential applications,
for which, in fact, it may be more useful to have higher oxygen vacancies and thus
more stable domain structures.

\begin{acknowledgments}
The authors thank J.-M. Triscone for useful discussions, and M. Lopes for technical support. Work at UniGE supported by the Swiss National Science Foundation through the NCCR MaNEP and Division II. Work at Yale supported by the National Science Foundation under MRSEC DMR-1119826, DMR-1006256, and FENA. ABK acknowledges the Universidad de Barcelona, Ministerio de Ciencia e Innovaci\'on (Spain) and  Generalitat de Catalunya for partial support through the I3 program.
\end{acknowledgments}

\bibliographystyle{prsty}

\end{document}